\documentclass[a4paper,12pt]{article}

\usepackage{graphicx}
\usepackage{array}
\usepackage{amsmath}
\usepackage{amssymb}
\usepackage{latexsym}
\usepackage{color}
\usepackage[normalem]{ulem}

\makeatletter
\@addtoreset{equation}{section}
\makeatother

\date{empty}
\begin{document}
\begin{titlepage}
\null
\begin{flushright}
%arXiv:YYMM.XXXX
TIT/HEP-631 \\
July, 2013
\end{flushright}
\vskip 0.5cm
\begin{center}
  {\Large \bf A Thermal Field Theory \\
\vskip 0.3cm
with Non-uniform Chemical Potential
}
\vskip 1.5cm
\normalsize
\renewcommand\thefootnote{\alph{footnote}}

{\large
Masato Arai$^{\dagger}$\footnote{masato.arai(at)fukushima-nct.ac.jp},
Yoshishige Kobayashi$^\ddagger $\footnote{yosh(at)th.phys.titech.ac.jp}
and Shin Sasaki$^\sharp$\footnote{shin-s(at)kitasato-u.ac.jp}
}
\vskip 1.0cm
  {\it
  $^\dagger$
  Fukushima National College of Technology \\ 
  Iwaki, Fukushima 970-8034, Japan \\
  \vskip 0.2cm
  $^\dagger$
  Institute of Experimental and Applied Physics, \\
  Czech Technical University in Prague, \\
  Horsk\' a 3a/22, 128 00, Prague 2, Czech Republic \\
  \vskip 0.2cm
  $^\ddagger$
  Department of Physics, Tokyo Institute of Technology \\
  Tokyo 152-8551, Japan \\
  \vskip 0.2cm 
  $^\sharp$
  Department of Physics,  Kitasato University \\
  Sagamihara 252-0373, Japan
}
\vskip 0.5cm
\begin{abstract}
We investigate thermal one-loop effective potentials 
in multi-flavor models with chemical potentials.
We study four-dimensional models in which each flavor 
has different global $U(1)$ charges.
Accordingly they have different chemical potentials.
We call these ``non-uniform chemical potentials,'' 
which are organized into a diagonal matrix $\hat{\mu}$.
The mass matrix at a vacuum does not commute with $\hat{\mu}$.
We find that the effective potential is divided into 
three parts. The first part is the Coleman-Weinberg potential.
The UV divergence resides only in this part.
The second is the correction to the Coleman-Weinberg potential
that is independent of temperature, and the third depends on both 
temperature and $\hat{\mu}$.
Our result is a generalization of the thermal potentials in 
previous studies for models with single and multi-flavors with
(uniform) chemical potentials
and reproduces all the known results correctly.
\end{abstract}
\end{center}

\end{titlepage}

\newpage
\setcounter{footnote}{0}
\renewcommand\thefootnote{\arabic{footnote}}
\pagenumbering{arabic}
%%%%%%%%%%%%%%%%%%%%%%%%%%%%%%%%%%%%%%%%%%%%%%%%%
\section{Introduction}
Thermal field theories with chemical potentials play an important
role to understand 
many issues in physics such as the quark-gluon plasma
\cite{Collins:1974ky, Cabibbo:1975ig}, 
the color superconductivity in QCD
\cite{Barrois:1977xd}, 
cosmology and astrophysics of neutron stars
\cite{Oppenheimer:1939ne}. 
One of characteristic features in these 
fields is phase transitions, which are possibly caused by quantum 
corrections, finite temperature and density effects.

When the theory admits perturbative approximations,
the one-loop effective potential
is a powerful tool to find the structure of vacua.
Thermal one-loop effective potentials of single flavor models with 
a chemical potential have been intensively studied
(see \cite{Ac1, Ac2, Qu} and references therein).
On the other hand, chemical potentials in multi-flavor models have been 
introduced by a flavor-independent way in the literature 
\cite{Kapusta:1981aa, Bernstein:1990kf, Benson:1991nj}. 
In general, each flavor can have a different value of chemical 
potential, which we call ``non-uniform chemical potentials.''
The non-uniform chemical potentials become 
important when one studies multi-flavor models with different $U(1)$
global charges such as generalized O'Raifeartaigh models \cite{Sh}
and so on. 

The generalized O'Raifeartaigh model has phenomenological interests since it 
yields a spontaneous supersymmetry breaking with $U(1)$ R-symmetry breaking, 
allowing gauginos to be massive.
The key issue to realize the $U(1)$ R-symmetry breaking is that the model
should include multi-flavors with a peculiar choice of different $U(1)$ R-charges \cite{Sh}.
The generalized O'Raifeartaigh models with finite temperatures have been 
studied to investigate thermal history of supersymmetry breaking vacua in 
the early universe
\cite{AbChJaKh1,CrFoWa,FiKaKrMaTo,AbJaKh,MoSc,Ka,KaPa, ArKoSa}.
Effects of finite temperature and a chemical potential to the 
$U(1)$ R-symmetry breaking have been 
studied for a supersymmetric model with a single flavor \cite{RiSe}
\footnote{Effects of non-uniform chemical potentials for lepton numbers are considered in the 
MSSM framework \cite{Bajc:1997ky, Bajc:1997wt, Bajc:1999he}. However, there 
are no mass matrices that do not commute with the chemical potentials.
This setting is essentially different from ours.}, 
where the chemical potential breaks the $U(1)$ R-symmetry even at high
temperatures. 
However, in this model a spontaneous supersymmetry breaking is not considered.
In order to study the thermal history of the generalized O'Raifeartaigh model, we need 
to understand how to calculate the effective potential with finite temperature and 
non-uniform chemical potential.

One reason that people have not paid attention to 
theories with non-uniform chemical potentials might be rather technical issue.
In a Lagrangian $\mathcal{L}$, the non-uniform chemical potentials 
are organized into 
 a diagonal matrix $\hat{\mu}$, which generically does not commute with 
a mass matrix $\hat{m}$ in $\mathcal{L}$.
As we will see in this paper, the straightforward generalization of 
formulas for single flavor models 
into those for models with the non-uniform chemical potential is not
valid 
due to the non-commutative nature of the matrices 
$\hat{\mu}$ and $\hat{m}$. 

The purpose of this paper is to establish the precise calculational
scheme of the thermal
one-loop effective potential of the multi-flavor models with 
non-uniform chemical potentials.
The calculations are quite delicate and need special attention on the UV
regularization due to the non-commutativity of the matrices $\hat{\mu}$
and $\hat{m}$. 
We find that the one-loop effective potential is the
sum of the Coleman-Weinberg potential \cite{CoWe}, which is independent 
of the temperature and $\hat{\mu}$, and the terms that depend on them.
Non-tirivial finding in this result is that the terms that depend on temperature 
and $\hat{\mu}$ are UV finite.
Since the main purpose of this paper is the calculations themselves, 
we will show the detail treatment of terms step by step in the
calculations. 

The organization of this paper is as follows.
In the next section, we review the calculation of the effective
potential in the single flavor models.
In section 3, we generalize the calculation to the multi-flavor models. 
This procedure involves various non-trivial aspects in the calculation.
In section 4, the result in section 3 is applied to a 2 flavors model
as a simple example.
Section 5 is devoted to conclusion and discussions. 
In appendix A, we prove the positive-definiteness of the matrix
$\sqrt{p^2 \mathbf{1} + \hat{m}^2} \pm \hat{\mu}$. 
In appendix B, we show the asymptotic behavior of the determinant
quantities defined in section 3.

\section{Effective potential -- single flavor model}
We begin with the model including
a single complex scalar field $\phi$.
The Lagrangian is 
\begin{eqnarray}
 \mathcal{L} = \partial_m \phi \partial^m \phi^\dagger- V(\phi,
\phi^{\dagger}), \label{pot}
\end{eqnarray}
where $m=0,1,2,3$ is 
the space-time vector index.
We use the mostly minus convention of the metric $\eta_{mn} =
\mathrm{diag} (1,-1,-1,-1)$.
The scalar potential $V(\phi, \phi^{\dagger})$ has at least one extrema
where the
scalar field $\phi$ develops its vacuum expectation value (VEV)
$\phi_{\mathrm{cl}}$.
The model exhibits a $U(1)$ global symmetry $\phi' =
e^{iq} \phi$ where $q$ is the $U(1)$ charge of $\phi$.
The one-loop effective potential of the model at finite temperature with 
a chemical potential is calculated through the partition function.
The temperature $T$ is introduced by letting 
$i x_0 = \tau$ and by imposing
the periodic boundary
condition $\phi(\tau,\vec{x})=\phi(\tau+\beta,\vec{x})$, where
$\beta=1/T$ and $\vec{x} = (x_1,x_2,x_3)$. The chemical potential
$\mu$ is introduced by gauging 
the $U(1)$ global symmetry in the Lagrangian (\ref{pot})
\cite{Ac1,Ac2, HaLaMu}.
The space-time derivative $\partial_m$ is replaced by the gauge
covariant derivative $D_m  = \partial_m  + i q A_m $.
The non-dynamical gauge field $A_m$ is introduced as a VEV only
in the zeroth
component $\langle A_m \rangle = (i \mu, {\bf 0})$.
Then the partition function is given as
\begin{eqnarray}
 Z={\rm Tr} e^{-\beta (H - \mu {\cal N})}=C \int_{\phi(\tau)=\phi(\tau+\beta)}
{\cal D}\phi {\cal D}\phi^\dagger 
 e^{-\int_0^\beta d\tau \int d^3x (D_0\phi D_0\phi^\dagger +
\vec{\nabla} \phi \cdot \vec{\nabla} \phi^\dagger
+V)},
\label{eq:partition_function}
\end{eqnarray}
where $H$ is the Hamiltonian associated with the Lagrangian
\eqref{pot} and ${\cal N}$ is the Noether current of the $U(1)$ 
symmetry. 
The constant $C$ is a normalization factor and 
$D_0={\partial \over \partial \tau}-\mu$.
Here we set $q = 1$ for simplicity.
$\vec{\nabla}$ is the differentiation with respect to $\vec{x}$.
The generating function is derived from the partition function
\eqref{eq:partition_function}
from which we obtain the Feynman rules 
including the temperature and the chemical potential.
With the use of the Feynman rules, the effective potential is
obtained as 
\cite{Ac1,Ac2}
\begin{eqnarray}
V_B^{\beta, \mu}
 (\phi_{\rm{cl}}) = V_B^{(0) \, \beta, \mu}(\phi_{\rm{cl}}) +
 V_B^{(1) \, \beta, \mu} (\phi_{\rm{cl}})
+ (\text{higher-loop\ corrections}).
\end{eqnarray}
The first term $V_B^{(0) \, \beta, \mu} (\phi_{\mathrm{cl}}) = V (\phi_{\mathrm{cl}},
\phi^{\dagger}_{\mathrm{cl}})$
is the potential at tree level. 
The second term $V_B^{(1) \beta, \mu}$ is the one-loop part
of the effective potential which is given by 
\begin{eqnarray}
& & V_B^{(1) \, \beta, \mu} (\phi_{\rm{cl}}) = - \frac{1}{2\beta} \sum_{n= - \infty}^{\infty} 
\int \! \frac{d^3 p}{(2\pi)^3} \log (\omega_n^2 + \omega_p^2),
\label{eq:eff_pot_single}
\\
& & 
\omega^2_p \equiv p^2 + m_B^2 , \quad 
\omega_n \equiv 2 \pi \beta^{-1} n - i \mu,
\quad 
n \in \mathbb{Z},
\nonumber 
\end{eqnarray}
where $m^2_B $ is the mass squared of $\phi$
at the vacuum $\phi=\phi_{\rm{cl}}$, 
\begin{equation}
m^2_B = \left. \frac{\partial^2 V}{\partial \phi^\dagger \partial \phi}
 \right|_{\phi=\phi_{\rm{cl}}}.
\end{equation}
The subscript ``$B$'' stands for quantities associated with boson fields.
In general, the summation over the discrete momentum modes $n$ diverges.
In order to regularize the infinity, we 
rewrite the summation over $n$ to an auxiliary
integration over $a^2 \in \mathbb{R}$ \cite{DoJa,Be}:
\begin{eqnarray}
\sum_{n=-\infty}^{\infty} 
\log (\omega^2_n + \omega_p^2)
= 
 \int^{\omega_p^2}_{1/\beta^2} \! da^2 
\sum_{n=-\infty}^{\infty} 
 \frac{1}{\omega_n^2 + a^2} 
+ \sum_{n=-\infty}^{\infty} 
\log (\omega_n^2 + 1/\beta^2). 
\label{int}
\end{eqnarray}
Next, we evaluate the sum $\sum_{n} \frac{1}{\omega_n^2 + a^2}$ 
in \eqref{int}. 
Since the function $\frac{\beta}{2} \cot
\left(\frac{\beta \omega}{2}\right)$ has poles at
$\omega = 2 \pi \beta^{-1} n $ with residue 1, 
the summation is rewritten as \cite{Be}
\begin{eqnarray}
\sum_{n = -\infty}^{\infty} \frac{1}{(2 \pi \beta^{-1} n - i \mu)^2 + a^2}
&=& 
\sum_{
\omega \in 2 \pi \beta^{-1} \mathbb{Z}
}
\frac{\beta}{2} \mathrm{Res} 
\left[
\cot
\left(
\frac{\beta \omega}{2} 
\right)
\frac{1}{(\omega - i \mu)^2 + a^2}
\right]
\nonumber \\
&=& - \sum_{\omega \not \in 2 \pi \beta^{-1} \mathbb{Z}}
\frac{\beta}{2} \mathrm{Res} 
\left[
\cot
\left(
\frac{\beta \omega}{2} 
\right)
\frac{1}{(\omega - i \mu)^2 + a^2}
\right],
\label{eq:sum_integration}
\end{eqnarray}
where the sum in the right hand sides is taken over poles of the
function in the square bracket.
Here we used the fact that 
the following contour integral vanishes for a sufficiently large 
circle $C_R$ with the radius $R$,
\begin{equation}
\lim_{R \to \infty} \oint_{C_R} \! d \omega \ \frac{\beta}{2} 
\cot 
\left(
\frac{\beta \omega}{2} 
\right)
\frac{1}{(\omega - i \mu)^2 + a^2} =0.
\label{a2-int}
\end{equation}
The poles of the function $[(\omega - i \mu)^2 + a^2]^{-1}$
are found to be $\omega = i (\mu \pm \sqrt{a^2})$.
Then the summation in the last expression in \eqref{eq:sum_integration}
is easily performed:
\begin{align}
& \sum_{\omega \not \in 2 \pi \beta^{-1} \mathbb{Z}}
\mathrm{Res} 
\left[
\cot
\left(
\frac{\beta \omega}{2} 
\right)
\frac{1}{(\omega - i \mu)^2 + a^2}
\right]
\nonumber \\
= & - \frac{1}{2 \sqrt{a^2}} 
\left[
\mathrm{coth}
\left(
\frac{\beta (\mu + \sqrt{a^2})}{2}
\right)
+
\mathrm{coth}
\left(
\frac{\beta (- \mu + \sqrt{a^2})}{2}
\right)
\right].
\label{eq:res_sum}
\end{align}
The above result is integrated by $a^2$ and
we obtain
\begin{align}
\sum_{n=-\infty}^{\infty} 
\log (\omega^2_n + \omega_p^2)
= 
 \log
\left[
\sinh
\left(
\frac{\beta}{2}( \omega_p + \mu)
\right)
\right]
+ 
 \log
\left[
\sinh
\left(
\frac{\beta}{2}( \omega_p - \mu)
\right)
\right]
+  C(T,\mu) ,
\label{eq:sum_log}
\end{align}
where $C(T,\mu)$ is a $p$-independent ``constant'', which may or may not depend
on $T$ and/or $\mu$. The ``constant'' $C (T,\mu)$
comes from the lower limit of the integration and
the infinite sum in \eqref{int}, namely,
\begin{equation}
C(T,\mu) =
 - \log \left[ \sinh \left( \frac{1}{2}( 1 + \beta \mu) \right) \right]
 - \log \left[ \sinh \left( \frac{1}{2}( 1 - \beta \mu) \right) \right]
+ \sum_{n=-\infty}^{\infty}  \log (\omega_n^2 + 1/\beta^2).
\label{eq:Diff-C}
\end{equation}

Actually $C(T,\mu)$ is a constant which is independent of both $T$ and $\mu$.
To see this fact, we differentiate $C(T,\mu)$
with respect to $\mu$:
\begin{equation}
\frac{\partial C(T,\mu)}{\partial \mu}
=
\frac{\beta}{2} \left[
- \coth \left( \frac{1}{2}( 1 + \beta \mu) \right)
+ \coth \left( \frac{1}{2}( 1 - \beta \mu) \right)
\right]
- \sum_{n=-\infty}^{\infty}
 \frac{2i \omega_n }{\omega_n^2 + 1/\beta^2}.
\label{eq:matsubara_sum}
\end{equation}

The convergence of the infinite sum in the above equation
is not obvious, however,
the sum can be deformed by using the symmetry $n \to -n$ of the sum,
\begin{eqnarray}
\sum_{n=-\infty}^{\infty}  \frac{2i \omega_n }{\omega_n^2 + 1/\beta^2}
 &=& \sum_{n=-\infty}^{\infty} \left[
\frac{1}{\beta^{-1} - i \omega_n} -\frac{1}{\beta^{-1} + i \omega_n}
\right]
\nonumber \\
&=& \sum_{n=-\infty}^{\infty} \left[
\frac{1}{2} \left(
\frac{1}{\beta^{-1} - i \omega_n} + \frac{1}{\beta^{-1} - i \omega_{-n}}
\right) -  \frac{1}{2} \left(
\frac{1}{\beta^{-1} + i \omega_n} + \frac{1}{\beta^{-1} - i \omega_{-n}}
\right) \right]
\nonumber \\
&=& \sum_{n=-\infty}^{\infty} \left[
  \frac{\beta^{-1} - \mu}{(\beta^{-1} - \mu)^2 +(2\pi\beta^{-1} n)^2}
- \frac{\beta^{-1} + \mu}{(\beta^{-1} + \mu)^2 +(2\pi\beta^{-1} n)^2}
\right] .
\end{eqnarray}
The infinite sum in the last line is convergent obviously.
The infinite sum can be calculated as shown previously,
which gives $\frac{\beta}{2} \coth \left( \frac{1}{2}( 1 - \beta \mu) \right)$
and $- \frac{\beta}{2} \coth \left( \frac{1}{2}( 1 + \beta \mu) \right)$.
Then we find
\begin{equation}
\frac{\partial C(T,\mu)}{\partial \mu}
= 0.
\end{equation}
Furthermore, in the case of $\mu=0$, $C (T,\mu)$ is independent of $T$\cite{Be}.
Thus $C (T,\mu)$ depends neither $T$ nor $\mu$.

Expanding the hyperbolic sine function and factoring out
$e^{\frac{\beta}{2} \omega_p}$ in \eqref{eq:matsubara_sum},
the one-loop part of the effective potential is given by 
\begin{eqnarray}
V_B^{(1) \, \beta, \mu}(\phi_{\mathrm{cl}}) = - \int \! \frac{d^3 p}{(2 \pi)^3} 
\left[
\frac{\omega_p}{2} + \frac{1}{2\beta} 
\log (1 - e^{- \beta (\omega_p
+ \mu)}) + 
\frac{1}{2\beta} 
\log (1 - e^{- \beta (\omega_p
- \mu)})
\right]
+ \mathrm{const},
\end{eqnarray}
where ``const.'' is just a numerical constant and does not contain the
VEV, $T$ and $\mu$.
The explicit value of the constant depends on models.
The first term in the integrand 
is the temperature-independent zero-point energy which
turns out to be the ordinary Coleman-Weinberg potential \cite{CoWe}. 
The second and third terms are contributions coming from finite
temperature and chemical potential.
We note that the potential becomes complex 
when $|\mu| > |m|$.
Thus the chemical potential has the upper bound $|\mu| \le |m|$.

It is straightforward to incorporate the fermionic contributions to the
effective potential. 
We consider the model with the following Lagrangian
\begin{align}
\mathcal{L} = \partial_m \phi \partial^m \phi^{\dagger} 
- V(\phi, \phi^{\dagger}) 
+ i \bar{\psi} \gamma^m \partial_m \psi 
- \bar{\psi} G (\phi, \phi^{\dagger}) \psi,
\label{eq:single_boson_fermion_L}
\end{align}
where $\psi$ is a Dirac fermion field and 
$G$ is a linear function of $\phi,
\phi^{\dagger}$. 
The fermion transforms as 
$\psi^\prime= e^{i\tilde{q}} \psi$
by the $U(1)$ group
where 
$\tilde{q}$
is the $U(1)$ charge of the fermion.
The function $G$ is chosen such that the Lagrangian
\eqref{eq:single_boson_fermion_L} is invariant under the $U(1)$
transformation and gives the mass for the fermion 
$m_F = G|_{\phi = \phi_{\mathrm{cl}}}$.
Now we promote the global $U(1)$ symmetry to gauge symmetry. 
The chemical potential is introduced as the VEV of the zeroth component
of the $U(1)$ gauge field, $\langle A_m \rangle = (i \mu, {\bf 0})$.
We calculate the one-loop contributions to the effective potential
from the fermion $\psi$.
Since the fermion satisfies the anti-periodic boundary condition along
the $\tau$ direction, the summation over the discrete momentum 
in \eqref{eq:sum_integration} is changed.
For the fermion field, the modes $n$ are shifted by $1/2$, namely,
$\omega_n = 2
\pi \beta^{-1} (n + 1/2) - i \mu$. 
Again we rewrite the summation 
$\sum_n \frac{1}{\omega_n^2 + a^2}$ 
into the contour integral by using the function $-\tan(\beta \omega/2)$
instead of
$\cot (\beta \omega/2)$.
The calculation is performed in parallel with
with the bosonic case. 
The result is 
\begin{eqnarray}
V_F^{(1) \, \beta, \mu} (\phi_{\mathrm{cl}})
= 2 \int \! \frac{d^3 p}{(2 \pi)^3} 
\left[
\frac{\omega_p }{2} + \frac{1}{2\beta} 
\log (1 + e^{- \beta (\omega_p
+ \mu)}) + 
\frac{1}{2\beta} 
\log (1 + e^{- \beta (\omega_p
- \mu)})
\right]
+ \mathrm{const}.,
\label{eq:fermion-eff-pot}
\end{eqnarray}
where ``const.'' is independent of the VEV, $T$ and $\mu$. 
The subscript ``$F$'' stands for quantities associated with fermion fields.
The factor 2 in front of the above integral appears
because the fermion $\psi$ has 
twice the degrees of freedom of a scalar field.
We also note that the sign in front of the exponential factor in the
logarithmic function is plus.
Therefore, in contrast to the bosonic case, 
the argument of the logarithmic function does not become
negative and there are no upper bound for the chemical potential
$|\mu|$ in the contribution from the fermionic field.

\section{Multi-flavor generalization}
In this section, we consider the multi-flavor models
including $N$ complex scalar fields $\phi^i \ (i=1, \cdots, N)$.
The Lagrangian is given by 
\begin{align}
\mathcal{L} = \partial_m \phi^{i} \partial^m
 \phi^{\dagger}_{i} - V(\phi^{i},
 \phi_{i}^{\dagger}).
\label{eq:multi_scalar_Lagrangian}
\end{align}
The model has a $U(1)$ global symmetry $\phi^{i \, \prime} = e^{i q^{i}}
\phi^{i}$ 
where $q^{i}$ are $U(1)$ charges of the fields $\phi^{i}$. 
In the multi-flavor case, 
one repeats similar steps discussed in the previous section
to obtain the effective potential.
In the single flavor model, it is easy to 
perform the summation over the modes $n$ through the $a^2$ integral as
in (\ref{a2-int}). 
However, we need to elaborate 
on the calculation in some points
in the multi-flavor models.
In the following subsections, 
we show how to perform the summation 
of the modes $n$ and discuss the regularization with respect to
the momentum integral.

\subsection{Effective potential}
The effective potential is a function of the 
VEVs of the scalar fields whose dependence is attributed to the mass
matrix of the fields. 
The $N \times N$ mass matrix at a vacuum is given by
\begin{equation}
(\hat{m}^2_B)_{ij}
= \left. \frac{\partial^2 V }{\partial\phi^{\dagger}_{i} \partial\phi^{j}} 
\right|_{\phi^{i} = \phi_{\mathrm{cl}}^{i}} .
\end{equation}
We assume that 
the matrix $(\hat{m}_B^2)^T = \hat{m}_B^2$ is real and symmetric
\footnote{
In general, $\hat{m}_B^2$ is a Hermitian matrix and
the off-diagonal blocks 
$\frac{\partial^2 V }{\partial\phi^{i} \partial\phi^{j}}$, 
$\frac{\partial^2 V }{\partial\phi^{\dagger}_{i}
\partial\phi^{\dagger}_{j}} $ 
appear in the mass matrix.
In this case, in order to make the mass matrix be real and symmetric
one may have to divide a complex field into two real fields, 
and rewrite the mass matrix in the bases of the real fields.
Although the apparent size of the mass matrix is doubled,
the subsequent discussion is essentially the same.
}%
.
In addition, we assume that 
the mass matrix $\hat{m}^2_B$ is positive definite
to avoid tachyonic modes at the vacuum.
As we explained in the previous section, 
the chemical potentials are introduced through the 
gauging of the $U(1)$ symmetry. We denote the 
chemical potential of $\phi^{i}$ as $\mu^{i} = q^{i} \mu$
where $\mu$ is a real parameter.
We define the $N \times N$ chemical potential matrix which is real
and diagonal,
\begin{eqnarray}
\hat{\mu} \equiv \mathrm{diag} (\mu^1, \cdots, \mu^{N}).
\end{eqnarray}
As in the single flavor case,
the one-loop part of the effective potential of the model
is found to be 
\begin{eqnarray}
V_B^{(1) \, \beta, \mu} (\phi^i_{\rm{cl}}) = -\frac{1}{2\beta}
\sum_{n= - \infty}^{\infty} 
\int \! \frac{d^3 p}{(2\pi)^3} \mathrm{Tr} \log
 (\omega_n^2 + \omega_p^2),
\label{eq:eff_pot}
\end{eqnarray}
where
$\phi^i_{\mathrm{cl}}$ is the VEV of $\phi^i$ and
\begin{equation}
\omega_p^2 \equiv p^2 \mathbf{1} + \hat{m}_B^2, \quad 
\omega_n \equiv 2 \pi \beta^{-1} n \mathbf{1} - i \hat{\mu}.
\end{equation}
Here $\mathbf{1}$ is the $N \times N$ unit matrix.
Now we calculate the summation over the discrete momentum modes $n$.
We define
\begin{eqnarray}
\hat{v} (\mu) \equiv \sum_{n= -\infty}^{\infty} \mathrm{Tr} \log 
\left[
p^2 \mathbf{1} + \hat{m}_B^2 + \omega_n^2 
\right].
\end{eqnarray}
Again the summation over $n$ diverges.
We generalize the single flavor result 
\eqref{eq:sum_integration} 
to the multi-flavor models.
In the multi-flavor case,
both the mass parameter and the chemical potential become matrices.
Following the single flavor case, we rewrite the sum over $n$ by the
$a^2 \in \mathbb{R}$ integral. Although $\omega_p$ and $\omega_n$ are matrices, we can
rewrite $\hat{v}(\mu)$ as 
\begin{eqnarray}
\hat{v} (\mu)
&=& 
\int^{p^2}_{0} \! d a^2 
\sum_{n = - \infty}^{\infty} 
\mathrm{Tr} 
\left\{
\left(
a^2 \mathbf{1} + \hat{m}_B^2 + \omega_n^2
\right)^{-1}\right\} + \mathrm{const.},
\label{eq:multi_integrand}
\end{eqnarray}
where the power $-1$ in the trace stands for the inverse matrix.
In the above equation, the last term ``$\mathrm{const.}$'' is not a
constant itself, because it may depend on $T$ and $\hat{\mu}$.
However, we can show the sum of the last term and
the value of the lower limit of the integration in the first term is
completely independent of both $T$ and $\hat{\mu}$,
as shown for $C(T,\mu)$ in \eqref{eq:sum_log}.
So hereafter we keep only the upper limit of the integration.
The summation over $n$ is written as  
\begin{eqnarray}
& & \sum_{n = -\infty}^{\infty} \mathrm{Tr}
\left\{\left(
a^2 \mathbf{1} + 
\hat{m}_B^2 
+ \omega_n^2
\right)^{-1}\right\}
\nonumber \\
& & 
\quad = - \sum_{\omega \not \in 2 \pi \beta^{-1} \mathbb{Z}}
\mathrm{Res}\left[ 
\frac{\beta}{2}
\cot 
\left(
\frac{\beta \omega}{2}
\right)
\mathrm{Tr}
\left\{
\left(
a^2 \mathbf{1} 
+ \hat{m}_B^2 + (\omega \mathbf{1} - i \hat{\mu})^2
\right)^{-1}
\right\}\right].
\label{eq:sum_res}
\end{eqnarray}

Before going to the situation
where $\hat{\mu} \not=0$, we first consider $\hat{\mu} = 0$ case.
Generally speaking finding the trace of the inverse matrix is
cumbersome,
however, when $\hat{\mu} = 0$, we can perform the calculation by
diagonalizing the matrix $\hat{m}^2_B$.
When $\hat{\mu} = 0$, the trace of the matrix is calculated as 
\begin{eqnarray}
&&\mathrm{Tr}
\left\{
\left(
\hat{m}^2_B + a^2 \mathbf{1} + \omega^2 \mathbf{1} 
\right)^{-1}
\right\} \nonumber \\
&& \qquad \qquad = \mathrm{Tr} \ 
\left\{
\mathrm{diag}
\left(
\frac{1}{a^2 + m_1^2 + \omega^2},
\frac{1}{a^2 + m_2^2 + \omega^2},
\cdots,
\frac{1}{a^2 + m_{N}^2 + \omega^2}
\right)\right\},
\end{eqnarray}
where we have used the fact that the 
trace of the inverse matrix is written as 
the sum of the inverse of the corresponding eigenvalues.
Here $m_i^2$ are the eigenvalues of the mass matrix $\hat{m}_B^2$.
Since the singularities of the integrand are located at $a^2 + m_i^2 +
\omega^2=0$,
we then find 
\begin{eqnarray}
\sum_{\omega \not \in 2 \pi \beta^{-1} \mathbb{Z}}
\mathrm{Res}\left[
\frac{\beta}{2}
\cot 
\left(
\frac{\beta \omega}{2}
\right)
\mathrm{Tr}
\left\{
\left(
a^2 \mathbf{1} 
+ \hat{m}^2_B + \omega^2 \mathbf{1}
\right)^{-1}
\right\}\right]
= - 
\sum_{i = 1}^{N}
\frac{\coth 
\left(
\frac{\beta}{2} \sqrt{a^2 + m_i^2}
\right)}{\sqrt{a^2 + m_i^2}}.
\end{eqnarray}
Finally, we perform the integration by $a^2$. 
The result in the case $\hat{\mu}=0$ is therefore 
\begin{eqnarray}
\hat{v} (0) = 
\int^{p^2}_{0} \! d a^2 \sum_{i = 1}^{N} 
\frac{\coth 
\left(
\frac{\beta}{2} \sqrt{a^2 + m_i^2}
\right)}{\sqrt{a^2 + m_i^2}}
= \frac{4}{\beta} \sum_{i=1}^{N} \log 
\left\{\sinh 
\left(
\frac{\beta}{2} \sqrt{p^2 + m_i^2} 
\right)
\right\}
+ \mathrm{const}.
\end{eqnarray}
This precisely recovers the thermal effective potential for 
the multi-flavor models \cite{Ac2}.

Now we consider the case 
$\hat{\mu} \not= 0$.
We need to evaluate the summation over the residues
\begin{eqnarray}
\sum_{\omega \not \in 2 \pi \beta^{-1} \mathbb{Z}}
\mathrm{Res}
\left[ 
\cot 
\left(
\frac{\beta \omega}{2}
\right)
\mathrm{Tr}
\left\{
\left(
a^2 \mathbf{1} + \hat{m}^2_B + (\omega \mathbf{1}- i \hat{\mu})^2
\right)^{-1}
\right\}\right].
\label{eq:trace}
\end{eqnarray}
When the matrices $\hat{m}_B$ and $\hat{\mu}$ can be simultaneously
diagonalized, the two matrices commute with each other 
$[\hat{m}_B,\hat{\mu}] = 0$ and the calculation is the same in
the previous section.
This is because, 
by using new diagonal matrices $\hat{m}'_B$ and $\hat{\mu}'$
which are obtained from $\hat{m}_B$ and $\hat{\mu}$,
the effective potential is the simple summation of
that of the $N$-independent single flavor (complex) fields.
In the following, we consider the general case where 
$[\hat{m}_B,\hat{\mu}] \neq 0$.

In order to evaluate the residues in \eqref{eq:trace}, we need to find
singularities of the function $\mathrm{Tr} (a^1 \mathbf{1} + \hat{m}^2_B
+ (\omega \mathbf{1} - i \hat{\mu})^2)^{-1}$.
In general, the inverse of a matrix $M$ is given by 
\begin{equation}
M^{-1} = (\det M)^{-1} \tilde{M},
\end{equation}
where $\tilde{M}$ is the cofactor matrix of $M$.
Since all the elements of the matrix $M = a^2 \mathbf{1} + \hat{m}^2_B + (\omega \mathbf{1}- i
\hat{\mu})^2$ are polynomials of $\omega$,
the elements of the cofactor matrix $\tilde{M}$ do not have any poles in $\omega$. 
Therefore, all the singularities in $\mathrm{Tr} M^{-1}$ come from the
zeros of $\det M$, namely, the singularities in the $\omega$-plane
satisfy the following equation,
\begin{equation}
\det M = \det (a^2 \mathbf{1} + \hat{m}_B^2 + (\omega \mathbf{1} - i 
\hat{\mu})^2) = 0.
\label{eq:det_zero}
\end{equation}
The determinant of $M$ is written as a polynomial of $\omega$, namely
\begin{eqnarray}
\det M = \prod_{i=1}^{2N} (\omega - \chi_i),
\end{eqnarray}
where $\chi_i$ are the solutions to the equation \eqref{eq:det_zero}.
Then we have
\begin{eqnarray}
& & \sum_{\omega \not \in 2 \pi \beta^{-1} \mathbb{Z}}
\mathrm{Res} 
\left[
\cot 
\left(
\frac{\beta \omega}{2}
\right)
\mathrm{Tr}
\left\{
\left(
a^2 \mathbf{1} + \hat{m}^2_B + (\omega \mathbf{1} - i \hat{\mu})^2
\right)^{-1}
\right\}
\right]
\nonumber \\
& & = \sum_{\{\omega| \det M = 0 \}}
\mathrm{Res} 
\left[
\cot
\left(
\frac{\beta \omega}{2}
\right)
\frac{1}
{
(\omega - \chi_1) (\omega - \chi_2) \cdots (\omega - \chi_{2N})
}
\mathrm{Tr} \tilde{M}
\right]
\nonumber \\
& & = \sum_{i=1}^{2N} 
\frac{
\cot
\left(
\frac{\beta \chi_i}{2}
\right)
}
{
(\chi_i - \chi_1) \cdots 
(\chi_i - \chi_{i-1}) (\chi_i - \chi_{i+1})
\cdots
(\chi_i - \chi_{2N})
}
\left.
\mathrm{Tr}
\tilde{M} 
\right|_{\omega = \chi_i}.
\label{eq:sum_over_n}
\end{eqnarray}
For simplicity, we assume that 
the solutions $\chi_i$ are general and not degenerate.
The next task is to perform the integration of \eqref{eq:sum_over_n} by
$a^2$.
In order to find the $a^2$ dependence of the solutions $\chi_i$, 
we show the following facts.
We consider $\det M$ as a function of $a^2$ and $\omega$,
$g (a^2, \omega) \equiv \det M$.
Then $\omega = \chi_i$ are solutions to the equation $g (a^2, \omega
(a^2)) = 0$. 
From the implicit function theorem, we have the
following relation
\begin{equation}
\frac{\partial \chi_i}{\partial a^2}
= - \left.
\frac{
\frac{\partial \det M}{\partial a^2}
}{
\frac{\partial \det M}{\partial \omega}
} \right|_{\omega = \chi_i} .
\label{eq:imp_th}
\end{equation}
The denominator in \eqref{eq:imp_th} is evaluated as 
\begin{equation}
\left. \frac{\partial \det M}{\partial \omega} \right|_{\omega = \chi_i} 
= 
(\chi_i - \chi_1) \cdots (\chi_i - \chi_{i-1}) (\chi_i - \chi_{i+1})
\cdots (\chi_i - \chi_{2N}).
\end{equation}
On the other hand, since $a^2$ enters into the each element in
the matrix $M$ as 
$M=a^2 {\bf 1}+\cdots$
where $\cdots$ are terms that are independent of $a^2$, 
the numerator in \eqref{eq:imp_th} is evaluated as 
\begin{eqnarray}
\frac{\partial \det M}{\partial a^2}
&=&
\det 
\left(
\begin{array}{cccc}
1 & 0 & \cdots & 0 \\
M_{21} & M_{22} & \cdots & M_{N2} \\
\vdots & \vdots & \ddots & \vdots\\
M_{N1} & \cdots & \cdots & M_{NN}
\end{array}
\right) + 
\det 
\left(
\begin{array}{cccc}
M_{11} & M_{12} & \cdots & M_{1N} \\
0 & 1 & \cdots & 0\\
\vdots & \vdots & \ddots & \vdots \\
M_{N1} & \cdots & \cdots & M_{NN}
\end{array}
\right) \nonumber \\
&&+\cdots +\det\left(
\begin{array}{cccc}
M_{11} & M_{12} & \cdots & M_{1N} \\
\vdots & \ddots & \vdots & \vdots \\
M_{(N-1)1} & M_{(N-1)2} & \cdots & M_{(N-1)N} \\
0 & 0 & \cdots & 1
\end{array}
\right).
\end{eqnarray}
This is nothing but the trace part of the cofactor
matrix. Then, we obtain
\begin{equation}
\frac{\partial \det M}{\partial a^2} = \mathrm{Tr} \tilde{M}.
\end{equation}
In short, we find the following formula
\begin{equation}
\frac{d \chi_i}{d a^2}
= - \frac{1}
{
(\chi_i - \chi_1) \cdots (\chi_i - \chi_{i-1}) (\chi_i - \chi_{i+1})
\cdots (\chi_i - \chi_{2N})
} 
\left. \mathrm{Tr} \tilde{M} \right|_{\omega = \chi_i}.
\end{equation}
Using this formula, we can perform the integration over $a^2$. The
result is 
\begin{eqnarray}
\int \! d a^2 \sum_{n = - \infty}^{\infty} 
\mathrm{Tr}
\left\{
\left(
a^2 \mathbf{1} + \hat{m}^2_B + 
\omega_n^2
\right)^{-1}
\right\}
= \frac{2}{\beta} \sum_{i = 1}^{2N} \log
\left(
\sin 
\left(
\frac{\beta \chi_i}{2}
\right)
\right) + \mathrm{const}.
\end{eqnarray}
Therefore, we obtain the following expression of the effective 
potential:
\begin{eqnarray}
V_B^{(1) \, \beta, \mu} (\phi_{\mathrm{cl}})
&=& - \frac{1}{2 \beta} 
\int \! \frac{d^3 p}{(2 \pi)^3} 
\sum_{i = 1}^{2N} \log
\left(
\sin 
\left(
\frac{\beta \chi_i}{2}
\right)
\right)
+ \mathrm{const}.
\label{eq:FD_pot_b}
\end{eqnarray}

Now we consider the model that includes $N$ Dirac fermionic
fields $\psi^i~(i=1,\cdots N)$,
\begin{align}
\mathcal{L} = \partial_m \phi^{i} \partial^m
 \phi^{\dagger}_{i} 
- V(\phi^{i}, \phi^{\dagger}_{i}) 
+ i \bar{\psi}_{i} \gamma^m \partial_m \psi^{i} 
- \bar{\psi}_{i} \hat{M}^{i} {}_{j} (\phi, \phi^{\dagger})
 \psi^{j}. 
\end{align}
The model has $U(1)$ global symmetry $\phi^{i \, \prime} = e^{i
q^{i}} \phi^{i}$ and $\psi^{i \, \prime} = e^{i \tilde{q}^{i}}
\psi^{i}$.
The same calculations are applied to the fermionic sector where the mass
$\hat{m}_B$ for the bosonic fields is 
replaced by that for the fermionic fields 
$\hat{m}_F = \hat{M}|_{\phi = \phi_{\mathrm{cl}}}$ 
and the summation is separated into the parts $n = 1,3,
\cdots$ and $n = 2,4, \cdots$ \cite{Qu}.
We just replace the function 
$(\beta/2) \cot (\beta \omega/2)$ with $- (\beta/2) \tan (\beta
\omega/2)$ in the calculation. 
Then the contributions to the one-loop effective potential from the
Dirac fermions $\psi^{i}$ are found to be 
\begin{eqnarray}
V_F^{(1) \, \beta, \mu}(\phi_{\mathrm{cl}})
&=& \frac{1}{2 \beta} 
\int \! \frac{d^3 p}{(2 \pi)^3} 
\sum_{i = 1}^{2N} \log
\left(
\cos 
\left(
\frac{\beta \chi_i}{2}
\right)
\right)
+ \mathrm{const}.
\label{eq:FD_pot_f}
\end{eqnarray}
The expressions \eqref{eq:FD_pot_b} and \eqref{eq:FD_pot_f} are
one of the main results in this work.
A few comments are in order.
The constant parts in \eqref{eq:FD_pot_b} and \eqref{eq:FD_pot_f}
are set to be zero which is determined in the limit $\hat{\mu} \to 0$.
The expression \eqref{eq:FD_pot_b} is the natural generalization of the
single flavor formula \cite{Ac1}.
We note that in order to find the explicit expression of the
effective potential, one needs to find
all the solutions $\chi_i$ to the equation $\det M = 0$. 
The detail of the solutions $\chi_i$ depends on the respective models
and 
we never look for the explicit solutions in this paper.
Although we have performed the summation over the modes $n$ and
regularized the
divergences coming from the sum, the expressions \eqref{eq:FD_pot_b} 
and \eqref{eq:FD_pot_f} still contain divergent part
stemming from the momentum integration.
In the following section, we study the properties of the solutions
$\chi_i$ that do not depend on the details of models.
We will show that the divergent part in the integration by the momentum
$p$ is isolated and regularized. 

\subsection{Properties of solutions}
\label{sec:Properties of solutions}
We have obtained the one-loop effective potentials 
\eqref{eq:FD_pot_b} and \eqref{eq:FD_pot_f}.
In order to find the explicit expressions of the potentials, one needs
the solutions to the equation for $\omega$:
\begin{align}
\det (p^2\mathbf{1} + \hat{m}^2 + (\omega \mathbf{1}- i \hat{\mu})^2)
= 0.
\label{eq:omega_eq}
\end{align}
Here we do not distinguish the bosonic and fermionic masses.
In the following, we study properties of the solutions to the
equation (\ref{eq:omega_eq}).
Hereafter we assume that $\hat{m}^2 - \hat{\mu}^2$ is a positive
definite matrix.
This condition is a generalization of the bound of the chemical
potential
$|\mu| \le m$ for a single bosonic field.
The violation of this condition 
implies the existence of tachyonic modes in the vacuum
for the scalar fields.

\paragraph{Pure imaginary nature of the solution}
Let $\omega$  be a solution to the equation \eqref{eq:omega_eq} and 
$|0 \rangle$ the eigenvector of the matrix $p^2\mathbf{1} + \hat{m}^2 +
(\omega\mathbf{1} - i
\hat{\mu})^2$ associated with the zero eigenvalue. Then we find
\begin{align}
\langle 0 | (p^2\mathbf{1} + \hat{m}^2 + (\omega \mathbf{1}- i
 \hat{\mu})^2) | 0 \rangle 
= \omega^2 + p^2 - 2 i \langle 0 | \hat{\mu} | 0 \rangle \omega + 
\langle 0 | (\hat{m}^2 - \hat{\mu}^2) | 0 \rangle 
= 0,
\label{eq:eqn-eigenvector}
\end{align}
where $\langle 0 |$ is the  Hermitian conjugate of $| 0 \rangle$, 
which is normalized as $\langle 0 | 0 \rangle =1$.
Therefore, we have 
\begin{align}
\omega = i 
\left(
\langle 0 | \hat{\mu} | 0 \rangle 
\pm
\sqrt{
p^2 + \langle 0 | (\hat{m}^2 - \hat{\mu}^2) |0 \rangle 
+ \langle 0 | \hat{\mu} | 0 \rangle^2
}
\right).
\label{eq:omega_imaginary}
\end{align}
Because \eqref{eq:eqn-eigenvector} is a necessary condition
for the given solution $\omega$ and the associated eigenvector $|0 
\rangle $,
\eqref{eq:omega_imaginary} does not completely determine $\omega$,
but only shows that $\omega$ is equal to one of the two quantities of
the right-hand side.
The quantity $\langle 0 | (\hat{m}^2 - \hat{\mu}^2) | 0 \rangle$ is
 positive
by the assumption, and $\langle 0 | \hat{\mu} | 0 \rangle$ is real
 because 
$\hat{\mu}$ is a real and diagonal matrix.
Therefore $\omega$ is always pure imaginary for any $p$. 
In the following discussions, we denote the imaginary solutions with 
the positive sign by $\chi_{i}^{+}$,  and with the negative sign by 
$\chi_{i}^{-}$.

\paragraph{Solutions and relation to their reference values}
For later use we clarify the relation between
the solutions to the equation \eqref{eq:omega_eq}
and  the eigenvalues of the matrices $ i(\hat{\mu} \pm
\sqrt{p^2 \mathbf{1} + \hat{m}^2})$.
We define the following function:
\begin{align}
f (\eta, \omega) =& 
\det 
\left[
\left(
\omega \mathbf{1} - i (\hat{\mu} - \sqrt{p^2 \mathbf{1} + \hat{m}^2}) 
\right)
\left(
\omega \mathbf{1} - i (\hat{\mu} + \sqrt{p^2 \mathbf{1} + \hat{m}^2})
\right)
+ 
\eta 
[\hat{\mu}, \sqrt{p^2 \mathbf{1} + \hat{m}^2}]
\right] 
\notag \\
= & \det 
\left[
p^2 \mathbf{1} + \hat{m}^2 - \hat{\mu}^2 - 2 i \hat{\mu} \omega +
 \omega^2 \mathbf{1} + 
(\eta - 1) [\hat{\mu}, \sqrt{p^2 \mathbf{1} + \hat{m}^2}]
\right], 
\label{eq:eta_equation}
\end{align}
where $0 \le \eta \le 1$ is a parameter.
The matrix $\sqrt{p^2 \mathbf{1} + \hat{m}^2}$ should be chosen
to be positive definite%
\footnote{
In general, there are $2^N$ square roots of an $N \times N$ matrix.
We have chosen the branch where all the eigenvalues of the matrix 
$\sqrt{p^2 \mathbf{1} + \hat{m}^2}$ are positive.
This is always possible when $\hat{m}^2$ is positive definite.
}
.
The equation \eqref{eq:omega_eq} is equivalent to $f (1,\omega) = 0$. 
We first solve $f (0,\omega) = 0$ and then discuss the properties
of the solutions to \eqref{eq:omega_eq}. 
When $\eta = 0$, the equation $f(0,\omega) = 0$
becomes
\begin{align}
f (0,\omega) = 
\det
\left(
\omega \mathbf{1} - i (\hat{\mu} - \sqrt{p^2 \mathbf{1} + \hat{m}^2})
\right)
\det
\left(
\omega \mathbf{1} - i (\hat{\mu} + \sqrt{p^2 \mathbf{1} + \hat{m}^2})
\right)
= 0.
\label{eq:f_eq}
\end{align}
The equation \eqref{eq:f_eq} is equivalent to 
\begin{align}
\det \left(
\omega \mathbf{1} - i (\hat{\mu} + \sqrt{p^2 \mathbf{1} + \hat{m}^2})
\right) = 0, 
\label{eq:det_factored_p}
\end{align}
or
\begin{align} 
\det \left(
\omega \mathbf{1} - i (\hat{\mu} - \sqrt{p^2 \mathbf{1} + \hat{m}^2})
\right) = 0.
\label{eq:det_factored_m}
\end{align}
The solutions to these equations 
are given by the eigenvalues of the matrices $M_{\pm} \equiv i 
(\hat{\mu} \pm
\sqrt{p^2 \mathbf{1} + \hat{m}^2})$ respectively.
Note that they are the exact solutions to
the equation $f(\eta,\omega) = 0$ 
as long as $[\hat{m},\hat{\mu}] = 0$.
Since $\sqrt{p^2 \mathbf{1} + \hat{m}^2} \pm \hat{\mu}$
are positive definite matrices (see Appendix \ref{pos}), 
the eigenvalues of $M_{+}$, which we call $\kappa^{+}_i$, are pure
imaginary with the
positive sign.
On the other hand, the eigenvalues of $M_{-}$, which we call 
$\kappa^{-}_i$,
are pure imaginary with the negative sign.

We find that the solutions to the equation 
$f(\eta,\omega) = 0$ are pure imaginary
in the whole range of $0 \le \eta \le 1$.
To see it, we rewrite the matrix in \eqref{eq:eta_equation} as
\begin{eqnarray}
&& p^2 \mathbf{1} + \hat{m}^2 - \hat{\mu}^2 - 2 i \hat{\mu} \omega +
 \omega^2 \mathbf{1} + 
(\eta - 1) \left[\hat{\mu}, \sqrt{p^2 \mathbf{1} + \hat{m}^2}\right]
 \nonumber \\
&& \quad = \eta \left(p^2 \mathbf{1} + \hat{m}^2 - \hat{\mu}^2 \right)
+(1-\eta) \left( \sqrt{p^2 \mathbf{1} + \hat{m}^2} - \hat{\mu} \right)
\left( \sqrt{p^2 \mathbf{1} + \hat{m}^2} +\hat{\mu} \right)
- 2 i \hat{\mu} \omega +
 \omega^2 \mathbf{1}.
 \nonumber \\
\end{eqnarray}
Now we consider the eigenvector $|0\rangle \neq 0$
 which satisfies the following
relation,
\begin{align}
\left[
\eta \left(p^2 \mathbf{1} + \hat{m}^2 - \hat{\mu}^2 \right)
+(1-\eta) \left( \sqrt{p^2 \mathbf{1} + \hat{m}^2} - \hat{\mu} \right)
\left( \sqrt{p^2 \mathbf{1} + \hat{m}^2} +\hat{\mu} \right)
- 2 i \hat{\mu} \omega +
 \omega^2 \mathbf{1}
\right]
| 0 \rangle
= 0.
\label{eq:zero-vector}
\end{align}
The solutions to $f(\eta, \omega) = 0$ are given by 
\begin{eqnarray}
&& \omega = i 
\langle 0 | \hat{\mu} | 0 \rangle 
 \nonumber \\
&& 
\pm i 
\sqrt{
\eta \left( p^2 
+ 
\langle 0 | 
(\hat{m}^2 - \hat{\mu}^2) 
| 0 \rangle \right )
+ 
(1-\eta) 
\langle 0 |
( \sqrt{p^2 \mathbf{1} + \hat{m}^2} - \hat{\mu} )
( \sqrt{p^2 \mathbf{1} + \hat{m}^2} +\hat{\mu} )
| 0 \rangle 
+
\langle 0 | \hat{\mu} | 0 \rangle^2
}.
\nonumber \\
\label{eq:omega_sol}
\end{eqnarray}
The matrix
$\left( \sqrt{p^2 \mathbf{1} + \hat{m}^2} - \hat{\mu} \right)
\left( \sqrt{p^2 \mathbf{1} + \hat{m}^2} +\hat{\mu} \right)$
is positive definite since both $\sqrt{p^2 \mathbf{1} + \hat{m}^2}
 - \hat{\mu}$
and $\sqrt{p^2 \mathbf{1} + \hat{m}^2} + \hat{\mu}$ are
real symmetric and positive definite matrices.
Therefore, the quantity inside the square root in \eqref{eq:omega_sol}
is positive.
Then the solutions $\omega$ to the equation $f(\eta,\omega) = 0$ 
are always pure imaginary and non-zero.
We note that the solutions $\omega$ 
change continuously when $\eta$ runs from $0$ to $1$. However, 
as we have seen in the above
discussion, the solutions $\omega$ in \eqref{eq:omega_sol} are always
pure imaginary and non-zero. 
Then $\kappa^{+}_i$ and $\kappa^{-}_i$ change continuously
into $\chi_i^+$ and $\chi_i^-$ respectively
without flipping their signs. 
Therefore we conclude that the solutions to the equation
\eqref{eq:omega_eq} 
are pure imaginary, a half of which have positive signs and
the other half have negative signs. 
All those have the one-to-one correspondence to $\kappa^{\pm}_i$.

\paragraph{Large-$p$ behavior}
In order to find the final expression of the effective potentials
\eqref{eq:FD_pot_b} and \eqref{eq:FD_pot_f}, we need to perform the
momentum integration.
The integral diverges at the large momentum $p$.
In the following, 
we examine the large-$p$ behavior of 
the solutions $\chi_i^{\pm}$.

The equation \eqref{eq:omega_eq} generally has
$2N$ solutions.
In the large-$p$ limit, the solutions can be written in 
the following form,
\begin{align}
\chi_i^{\pm} = i \left( \mu_{i} \pm \sqrt{p^2 + m_{ii}^2}+\mathcal{O}(1/p)
 \right)
= i \left( \pm p + \mu_i +\mathcal{O}(1/p) \right),
\label{eq:appxsol-largeP}
\end{align}
where $\mu^i$ is the $i$th-diagonal component of $\hat{\mu}$ and
$m_{ii}^2$ is a diagonal $(i,i)$-component of $\hat{m}^2$.
Although all $\omega$ are linearly divergent at large-$p$, 
the solutions are distinguishable from each other at order 
$\mathcal{O}(p^0)$
as long as the chemical potentials $\mu_i$ are general values.
The eigenvalues $\kappa^{\pm}_i$ have the same asymptotic form as in
 \eqref{eq:appxsol-largeP}.
In this sense,
$\kappa^{\pm}_i$ are the approximate solutions to the equation
\eqref{eq:omega_eq} at least in the large-$p$ regime.
Since all the solutions $\chi^{\pm}_i$ to the equation
\eqref{eq:omega_eq} ($f(1,\omega) =0$) have the one-to-one
correspondence
to the solutions $\kappa^{\pm}_i$ to $f(0,\omega) = 0$,
each $\chi^{+}_i$ ($\chi^{-}_i$) approaches the corresponding one among 
the set $ \{ \kappa^{+}_i\}$ ($ \{ \kappa^{-}_i\}$).
In the following, we estimate
how $\chi^{\pm}_i$ approach their counterpart. 

The determinant of \eqref{eq:omega_eq} is rewritten as
\begin{eqnarray}
\det (p^2 \mathbf{1} + \hat{m}^2 + (\omega \mathbf{1} - i \hat{\mu})^2) 
=
\prod_{i=1}^{N} (\omega-\chi^+_i)(\omega-\chi^-_i)
=\prod_{i=1}^{N} \Delta^+_i \Delta^-_i,
\label{eq:DetProd}
\end{eqnarray}
where 
$\Delta_i^\pm \equiv \omega - \chi^\pm_i$.
Differentiating the both sides of \eqref{eq:DetProd} 
with respect to $\omega$, we obtain
\begin{align}
&\frac{d}{d \omega}
\det (p^2 \mathbf{1} + \hat{m}^2 + (\omega \mathbf{1} -i \hat{\mu})^2) 
\notag \\
&
\quad =\left( \Delta^+_2 \cdots \Delta^+_N \Delta^-_1 \cdots \Delta^-_N 
\right)
+ \left( \Delta^+_1 \Delta^+_3 \cdots \Delta^+_N \Delta^-_1 \cdots 
\Delta^-_N
\right)
 \notag \\
& \quad \quad +\cdots
+ \left(\Delta^+_1 \cdots \Delta^+_N \Delta^-_1 \cdots \Delta^-_{N-2}
 \Delta^-_{N}
\right)
+ \left( \Delta^+_1 \cdots \Delta^+_N \Delta^-_1 \cdots \Delta^-_{N-1}
 \right).
\label{eq:DofDetProd}
\end{align}
Now we choose one of $\{ \kappa^{\pm}_i \}$, say $\kappa^+_k$,
and substitute $\omega$ with $\kappa^+_k$.
As is shown in appendix \ref{large},
in the large-$p$ limit, \eqref{eq:DetProd} and \eqref{eq:DofDetProd}
behave in a way such as
$\det (p^2 \mathbf{1} + \hat{m}^2 + (\kappa^+_k \mathbf{1}
-i \hat{\mu})^2) \sim \mathcal{O} (p^{N-4})$ and
$\frac{d}{d \omega} \det (p^2 \mathbf{1} + \hat{m}^2 +
(\kappa^+_k \mathbf{1} -i \hat{\mu})^2) \sim \mathcal{O} (p^N) $.
We introduce the parameters $\varepsilon^\pm_i$ that 
represents the power of $p$ in
$\Delta^\pm_i$ in the large-$p$, namely 
we have $\Delta^\pm_i \sim \mathcal{O} (p^{\varepsilon^\pm_i})$.
The asymptotic behavior of \eqref{eq:DetProd} and \eqref{eq:DofDetProd}
tells us that
\begin{eqnarray}
&S = N-4 \qquad \left(S \equiv \displaystyle\sum_{i=1}^N (\varepsilon^+_i
 + \varepsilon^-_i) \right),
 & \label{eq:cond1a-forepsilon} \\
&\displaystyle \max_{i =1, \cdots, N}
\left\{ S - \varepsilon^+_i, S - \varepsilon^-_i \right\} = N
.
\label{eq:cond1b-forepsilon}
\end{eqnarray}
On the other hand, because $\kappa^+_k$ has the positive sign,
the large-$p$ behavior of each $\Delta^\pm_i$ in the case that $\omega
 =\kappa^+_k $
is given by
\begin{equation}
\Delta^+_i = \mathcal{O}(p^0), \quad \Delta^-_i = +2ip + \mathcal{O}(p^0)
\qquad (i=1,2,\cdots,N),
\end{equation}
from the asymptotic behavior \eqref{eq:appxsol-largeP}.
So one can impose the following conditions on $\varepsilon^\pm_i$ ,
\begin{align}
\varepsilon^-_1 = \varepsilon^-_2 = \cdots = \varepsilon^-_N = 1,
\label{eq:cond2-forepsilon}
\end{align}
and
\begin{align}
\varepsilon^+_i \le 0 \quad (i = 1, \cdots , N).
\label{eq:cond3-forepsilon}
\end{align}
The only possibility satisfying all the conditions
 \eqref{eq:cond1a-forepsilon}, \eqref{eq:cond1b-forepsilon},
 \eqref{eq:cond2-forepsilon} and \eqref{eq:cond3-forepsilon} is
\begin{align}
\varepsilon^+_i =
\left\{ \begin{array}{ll}
 -4 & ( i = l ) \\
 0 & ( i \neq l )
\end{array} \right. {\rm{for\ some\ }} l.
\end{align}
Among $2N$ components of $\Delta^\pm_i$,
only $\Delta^+_l=\chi^+_l-\kappa^+_k$ behaves $\Delta^+_l 
=\mathcal{O}(1/p^4)$.
It is obvious that $\chi^+_l$ is what is the corresponding solution to
 $\kappa^+_k$.
Though both $\chi^+_l$ and $\kappa^+_k$ diverge in the large-$p$ limit,
the difference $\chi^+_l - \kappa^+_k$ decreases at the inverse of
fourth power of $p$.
The above discussion can be applied to arbitrary $\kappa^\pm_i$.
Therefore one can conclude that the difference between any $\chi^\pm_l$
 and
the corresponding  $\kappa^\pm_k$
drops as $\mathcal{O}(1/p^4)$ in $p \to \infty$\footnote{
Indeed, the sum over all the differences $\chi_i^{\pm} - \kappa_i^{\pm}$
drops more quickly,
\begin{align}
\sum_{i=1}^N \chi^\pm_i = \sum_{i=1}^N \kappa^\pm_i + \mathcal{O}(1/p^5). 
\nonumber
\end{align}
However the $\mathcal{O} (p^{-4})$ behavior of $\Delta_i^{\pm}$ at 
large-$p$
is sufficient to discuss
the convergence of the integration over the momentum $\vec{p}$.
}.

\subsection{Regularized effective potential}
In this subsection, we study the regularization of the momentum
integral in the effective potentials \eqref{eq:FD_pot_b} and
\eqref{eq:FD_pot_f}.
Since
the solutions $\chi^{\pm}_i$ are pure imaginary, these are
represented as $\chi^{\pm}_i = \pm i |\chi^{\pm}_i|$.
We rewrite the potential \eqref{eq:FD_pot_b} in such a way that the
logarithmic function is well-defined in the large-$p$ regime:
\begin{align}
\log \left( \sin \frac{\beta}{2} \chi^\pm_i \right) &=
\log \left( \pm \sin \frac{i \beta}{2} |\chi^{\pm}_i| \right) \notag \\
 & = \log \left( \pm \frac{i}{2}
 e^{\frac{\beta}{2} |\chi^{\pm}_i|} ( 1- e^{- \beta |\chi^{\pm}_i|} ) \right) \notag \\
 &= \mp i \frac{\beta}{2} \chi^\pm_i
 + \log \left( 1 - e^{-\beta | \chi^\pm_i |} \right) + \log (\pm i/2).
\label{eq:decomposition-log-for-scalar}
\end{align}
The last term is the constant that includes the phase factor of 
$\chi_i^{\pm}$.
Then the sum of all $2N$ solutions becomes
\begin{align}
& \sum_{i=1}^{N} \left[
 \log \left( \sin \left( \frac{\beta \chi^+_i}{2} \right) \right)
+\log \left( \sin \left( \frac{\beta \chi^-_i}{2} \right) \right) 
\right]
\notag \\
& \quad = - i \frac{\beta}{2} \sum_{i=1}^{N} ( \chi^+_i - \chi^-_i ) 
+ \sum_{i=1}^{N} \left[ \log \left( 1  - e^{-\beta|\chi^+_i|} \right)
+ \log \left( 1  - e^{-\beta|\chi^-_i|} \right)
\right].
\label{eq:SumofLog}
\end{align}
The first term in the last line contains the Coleman-Weinberg potential.
Here we drop an irrelevant constant in the last line.
The sum of all the phase factors vanishes because the sum contains
$\log i$ as many as $\log(-i)$.

In order to find the large-$p$ behavior of the quantity
\eqref{eq:SumofLog}, we define 
$\delta^\pm_i(p) \equiv \chi^\pm_i(p) - \kappa^\pm_i(p) $.
Then the first term in \eqref{eq:SumofLog} is split into two parts,
\begin{eqnarray}
\sum_{i=1}^N ( \chi^+_i - \chi^-_i ) &=& 
\sum_{i=1}^N \left\{(\kappa^+_i +\delta^+_i) - (\kappa^-_i +\delta^-_i)
\right\}
\nonumber \\
&=& 2 i \ 
\sum_{i=1}^N \sqrt{p^2 + m^2_{ii}}
 + \sum_{i=1}^N (\delta^+_i - \delta^-_i).
\label{eq:SumofChi}
\end{eqnarray}
Here we used the relation 
$\displaystyle \sum_{i=1}^{N} \kappa_i^{\pm} = i\; \mathrm{Tr} \ 
(\hat{\mu} \pm
\sqrt{p^2 \mathbf{1} + \hat{m}^2})$.
Collecting all the terms
together, we obtain the one-loop part of the effective potential as
\begin{eqnarray}
 V_B^{(1) \, \beta,\mu}&=& - {1 \over \beta}
  \int {d^3p \over (2\pi)^3} \ \sum_{i=1}^N 
     \Big[
       \beta \sqrt{p^2+m_{Bii}^2}
       -{i\beta \over 2}\left((\chi^+_i - \kappa^+_i)
       - (\chi^-_i - \kappa^-_i)\right)
 \nonumber \\
&& \qquad
       +\log \left( 1  - e^{-\beta|\chi^+_i|} \right)
       + \log \left( 1  - e^{-\beta|\chi^-_i|} \right)
     \Big].
\label{eq:final_result}
\end{eqnarray}
The first term 
in the integrand is nothing but
the ordinary Coleman-Weinberg potential term,
of which regularization scheme is well known \cite{CoWe}.
The second term is corrections 
to the Coleman-Weinberg potential
stemming from the non-uniform chemical potential.
Because $\chi^\pm_i - \kappa^\pm_i$ decreases like $p^{-4}$ at large-$p$,
this term stays finite after the integration over the momentum.
The third and fourth terms are corrections from both chemical
potentials and temperature.
They are the familiar bosonic thermal potentials
where the argument of the
exponential functions are replaced by $\chi^\pm_i$.
The momentum integration of these terms converges since the exponential
factors behave well in the large-$p$ regime.
Therefore the divergent part comes only from the Coleman-Weinberg
potential term.
We stress that in the derivation of the first term in \eqref{eq:final_result}, 
the contributions coming from the chemical potential $\hat{\mu}$ are canceled out.
We emphasize that all the divergent pieces in \eqref{eq:final_result}
are completly independent of the temperature and the chemical potential. 
This is the necessary condition for the renormalization of the quantum
field theory.

\subsection{
Upper bound of $\hat{\mu}$ in fermionic contributions
}
As in
the case of the single-flavor models, it is 
straightforward to derive the contribution from fermions to the
one-loop effective potential
in the multi-flavor models.
However, there is one significant difference
on the upper bound of the chemical potentials.
So far we have assumed that $\hat{m}^2 - \hat{\mu}^2$ is
positive definite.
For the chemical potentials of the scalar fields, 
the upper bound has clear physical meaning.
Actually it is equivalent to the condition that the system has
no tachyonic modes at tree-level.
However the physical reason of the upper bound on the chemical potential
is not obvious for fermionic fields.
Indeed, the contributions to the one-loop thermal effective potential 
from the fermion in the single flavor model
\eqref{eq:fermion-eff-pot}
is well-defined both $\mu < m$ and $\mu \ge m$.
For the multi-flavor models, the positive definiteness of 
the matrix $\hat{m}^2 - \hat{\mu}^2$ 
leads to the property that the solutions to \eqref{eq:omega_eq},
$\omega$, are pure imaginary
and are divided into equal number of solutions $\chi^+_i$ and $\chi^-_i$.
These properties are not guaranteed if 
we do not assume the positive definiteness of the matrix.

We take a closer look at the potential \eqref{eq:FD_pot_f}.
When $\omega$ is pure imaginary, namely $\omega = i r \ (r \in
 \mathbb{R})$,
the integrand of \eqref{eq:FD_pot_f}
can be decomposed as,
\begin{align}
\log \left( \cos \frac{\beta}{2} \omega \right) &=
\log \left( \cos \frac{i \beta}{2} r \right) \notag \\
 & = \log \left( \frac{1}{2} e^{\frac{\beta}{2} r } ( 1+ e^{- \beta r} ) 
\right) \notag \\ 
 &= \frac{\beta}{2} \rm{Im} (\omega)
 + \log \left( 1 + e^{-\beta \rm{Im}(\omega)} \right) - \log 2.
\label{eq:decomposition-log-for-fermion}
\end{align}
In \eqref{eq:decomposition-log-for-fermion}, 
the last term is real and there are
no complex phase terms (see the last term in
\eqref{eq:decomposition-log-for-scalar} for the scalar field 
contributions)
and the point $\omega=0$ is not singular.
Therefore for fermionic fields, unlike for scalar fields,
the potential is well-defined and physical 
as long as $\omega$ is pure imaginary, and $\omega$ can freely move on
the imaginary axis including zero.
As we have shown, $\omega$ is pure imaginary when the following
quantity is positive:
\begin{equation}
p^2 + \langle 0 | (\hat{m}^2 - \hat{\mu}^2) |0 \rangle 
+ \langle 0 | \hat{\mu} | 0 \rangle^2.
\label{eq:discriminant}
\end{equation}
In the single-flavor models, \eqref{eq:discriminant} becomes $p^2 + m^2$
which is real and positive and therefore $\omega$ is always pure 
imaginary.
It is the reason that the potential is well-defined
both $\mu < m$ and $\mu \ge m$.
In the multi-flavor models, it becomes a little more complicated.
When $\hat{m}^2 - \hat{\mu}^2$ is positive definite, the solutions
$\omega$ are pure imaginary.
However when $\hat{m}^2 - \hat{\mu}^2$ is not positive definite,
it can be negative at small $p$.
In that case, $\omega$ may have not only an imaginary part but
a real part.
Then complex quantities may come out
from the logarithmic function in the effective potential,
which cannot be absorbed into a $p$-independent constant.
As a result the potential becomes complex value, which may be
recognized as the instability of the corresponding vacuum.
This is a new phenomenon only for 
multi-flavor models with non-uniform chemical potential.
In the case that $\hat{m}^2 - \hat{\mu}^2$ is not positive definite,
the small $p$ behaviors of the solutions $\omega$
depend on the details of the matrices $\hat{m}$ and $\hat{\mu}$.
We did not find any restriction that ensures the appropriate behaviors
of $\omega$ at small $p$.
Even if all $\omega$ are pure imaginary for the entire range of $p$,
each $\omega$ does not have the definite sign any more,
which may flip as $p$ goes from zero to infinity.
Therefore in the cases where $\hat{m}^2 - \hat{\mu}^2$ is not positive
definite, in order to extract the UV divergent piece as in
\eqref{eq:decomposition-log-for-fermion}, 
we need to extract the linear term in $\omega$ with the appropriate 
sign.
This sign is determined by its large-$p$ behavior 
which is necessary for the convergence of the momentum integral of the
logarithmic function.

With the above discussion,
the one-loop part of the effective potential from fermionic fields is given by
\begin{eqnarray}
 V_F^{(1) \, \beta,\mu}&=&{2 \over \beta}
  \int {d^3p \over (2\pi)^3} \sum_{i=1}^N 
     \Big[
       \beta \sqrt{p^2+m_{Fii}^2}
       -{i\beta \over 2}\left( (\chi^{\prime +}_i - \kappa^{\prime +}_i)
       - (\chi^{\prime -}_i - \kappa^{\prime -}_i) \right)
 \nonumber \\
&& \qquad
       +\log \left( 1 + e^{-\beta|\chi^{\prime +}_i|} \right)
       + \log \left( 1 + e^{-\beta|\chi^{\prime -}_i|} \right)
     \Big],
\label{eq:final_result_fermion}
\end{eqnarray}
where we have denoted the solutions to $f(1,\omega) = 0$ and 
$f(0,\omega) = 0$ as $\chi^{\prime \pm}_i$ and $\kappa^{\prime \pm}_i$
respectively.
Here the solutions $\chi^{\prime +}_i$ ($\chi^{\prime -}_i$) and
$\kappa^{\prime +}_i$ ($\kappa^{\prime -}_i$) have positive (negative)
sign at large-$p$.

\section{Simple example}
In this section,
we demonstrate an example of the explicit calculation of the thermal one-loop
potential \eqref{eq:final_result}. 
We consider a two-flavor bosonic model as the most simple multi-flavor model.
The mass matrix and chemical potential of the model are
\begin{equation}
\hat{m}^2= m^2 \left(
\begin{array}{cc}
 5 & -4 \\
-4 &  5 
\end{array}
\right), \qquad
\hat{\mu}= \mu \left(
\begin{array}{cc}
 1 & 0 \\
 0 &  -1 
\end{array}
\right),
\label{eq:matrix-2-flavor}
\end{equation}
where $m > \mu > 0$ are real parameters. 

The matrices $\hat{m}$ and $\hat{\mu}$ do not commute with each other 
and $\hat{m}^2 - \hat{\mu}^2$ is positive definite as long as
$m > \mu$. 
The two eigenvalues of the mass matrix $\hat{m}^2$ are $m^2$ and $9 m^2$.
The square root of the matrix $p^2 \mathbf{1}+ \hat{m}^2$ can be written 
in an explicit form by
\begin{equation}
\sqrt{p^2 \mathbf{1} + \hat{m}^2}= \frac{1}{2} \left(
\begin{array}{cc}
 \sqrt{p^2+m^2} + \sqrt{p^2+9m^2}  &  \sqrt{p^2+m^2} - \sqrt{p^2+9m^2} \\
 \sqrt{p^2+m^2} - \sqrt{p^2+9m^2}  &  \sqrt{p^2+m^2} + \sqrt{p^2+9m^2}
\end{array}
\right).
\end{equation}
The eigenvalues of $i (\hat{\mu} + \sqrt{p^2 \mathbf{1} + \hat{m}^2}) $ 
and $i (\hat{\mu} - \sqrt{p^2 \mathbf{1} + \hat{m}^2}) $ are
imaginary with positive sign and negative sign, respectively.
We assign those eigenvalues to $\kappa_i^+$ and  $\kappa_i^-$ as
\begin{eqnarray}
\kappa_1^+ = - \kappa_1^- &=&  
\frac{i}{2} \left(
\sqrt{p^2+m^2} + \sqrt{p^2+9m^2} +
\sqrt{\left( \sqrt{p^2+m^2} - \sqrt{p^2+9m^2} \right)^2 + 4\mu^2}
\right), \\
\kappa_2^+ = - \kappa_2^- &=&  
\frac{i}{2} \left(
\sqrt{p^2+m^2} + \sqrt{p^2+9m^2} -
\sqrt{\left( \sqrt{p^2+m^2} - \sqrt{p^2+9m^2} \right)^2 + 4\mu^2}
\right).
\end{eqnarray}

The equation \eqref{eq:omega_eq} is
\begin{align}
 & \det (p^2\mathbf{1} + \hat{m}^2 + (\omega \mathbf{1}- i \hat{\mu})^2)
\nonumber \\
&  \qquad = \omega^4 +2(p^2+5m^2+\mu^2) \omega^2 + p^4 +(10 m^2 - 2 \mu^2) p^2
 + 9m^4 - 10 \mu^2 m^2+ \mu^4 \nonumber \\
& \qquad = 0.
\end{align}
This equation 
has four solutions which can be found analytically,
\begin{eqnarray}
\chi^+_1 &=& i \sqrt{p^2 +5m^2 +\mu^2 + 2 \sqrt{\mu^2 p^2 + 4 m^4 + 5 \mu^2 m^2} },
 \\
\chi_2^+ &=& i \sqrt{p^2 +5m^2 +\mu^2 - 2 \sqrt{\mu^2 p^2 + 4 m^4 + 5 \mu^2 m^2} },
 \\
\chi_1^- &=& -i \sqrt{p^2 +5m^2 +\mu^2 + 2\sqrt{\mu^2 p^2 + 4 m^4 + 5 \mu^2 m^2} },
 \\
\chi_2^- &=& -i \sqrt{p^2 +5m^2 +\mu^2 -2 \sqrt{\mu^2 p^2 + 4 m^4 + 5 \mu^2 m^2} }.
\end{eqnarray}
The sum of $\kappa^\pm_i$ is
\begin{equation}
\frac{i}{2} \sum_{i=1}^2 \left( \kappa^+_i - \kappa^-_i \right)=
\sqrt{p^2+m^2} + \sqrt{p^2+9m^2},
\end{equation}
which gives the Coleman-Weinberg potential after applying
an appropriate regularization.
In the large $p$ limit, the difference between coresponding $\kappa^\pm_i$
and $\chi^\pm_i$ is expanded in the series of $1/p$,
\begin{eqnarray}
\chi^+_1 - \kappa^+_1 &=& -(\chi^-_1 - \kappa^-_1) =
- 2i \frac{\mu m^4}{p^4} + 2 i \frac{\mu^2 m^4}{p^5}+\mathcal{O}(p^{-6}), \\
\chi^+_2 - \kappa^+_2 &=& -(\chi^-_2 - \kappa^-_2) =
 2i \frac{\mu m^4}{p^4} + 2 i \frac{\mu^2 m^4}{p^5}+\mathcal{O}(p^{-6}),
\end{eqnarray}
which decrease at the order of $\mathcal{O}(p^{-4})$
as proved in the section \ref{sec:Properties of solutions}.
The sum of them, on the other hand, behaves like the order 
$\mathcal{O}(p^{-5})$
\begin{equation}
\sum_{i=1}^{2} \chi^+_i - \kappa^+_i
= 4  i \frac{\mu^2 m^4}{p^5}+\mathcal{O}(p^{-6}),
\end{equation}
as mentioned in the footnote in the section \ref{sec:Properties of solutions}.
Finally we obtain the thermal effective potential of this example,
\begin{eqnarray}
V_B^{(1) \, \beta,\mu} &=&  - \int {d^3p \over (2\pi)^3} \left[
\sqrt{p^2+m^2} + \sqrt{p^2+9m^2} \Big. \right.
+ \frac{i}{2} \sum_{i=1}^2
\left( \chi_i^+ - \chi_i^- - \kappa_i^+  + \kappa_i^- \right)
 \nonumber \\
 && \left.
+ \frac{1}{\beta} \sum_{i=1}^2 \left(
 \log \left( 1 - e^{i \beta \chi_i^+} \right)
+ \log \left( 1 - e^{-i \beta \chi_i^-} \right)
\right)
 \right].
\end{eqnarray}

\section{Conclusion and discussions}
In this paper, we studied the thermal one-loop effective potential of
multi-flavor models with the non-uniform chemical potential.
The non-uniform chemical potential is 
necessary when the fields in the multi-flavor models have different 
$U(1)$ charges. 

Due to the non-commutativity between the matrices $\hat{\mu}$ and 
$\hat{m}$, the rigorous treatment of the non-uniform chemical
potentials in the effective potential had
been less understood in the literature.
We wrote down the explicit expressions of the thermal effective
potentials with the non-uniform chemical potential.
We showed the detail calculations 
in each step and obtained the expressions \eqref{eq:FD_pot_b} and
\eqref{eq:FD_pot_f}. 
The effective potential is completely determined by the
zeros of the determinant $\det (p^2 \mathbf{1} + \hat{m}^2 + 
(\omega \mathbf{1}- i
\hat{\mu})^2)$, which is the natural generalization of
the single flavor case. 
Although the expressions \eqref{eq:FD_pot_b} and \eqref{eq:FD_pot_f} 
contain terms with UV
divergent piece, the careful analysis of the large-$p$ behavior of the
solution to $\det (p^2 \mathbf{1} + \hat{m}^2 + (\omega \mathbf{1}- i
\hat{\mu})^2) = 0$ reveals that the divergent parts of the effective
potential come only from the Coleman-Weinberg potential. 
At the same time, we found that terms that 
depend on temperature and the chemical potential give finite 
contributions to the effective potential.
The generalization to the models with fermions is straightforward
except subtleties of the sign of the solutions.
We then obtained the final result \eqref{eq:final_result} and
\eqref{eq:final_result_fermion}.
In supersymmetric models, the Coleman-Weinberg potential
is exactly canceled out because the bosonic and fermionic mass matrices
are coincident. 
In the case of models with non-uniform chemical potentials, 
however, there are finite corrections to the Coleman-Weinberg potential
that depends on $[\hat{m},\hat{\mu}]$, which are not canceled 
in general.

We again stress that the result \eqref{eq:final_result} is not the
simple generalization of the well-known single flavor case. 
The non-commutativity of the chemical potential matrix $\hat{\mu}$ and
the mass matrix $\hat{m}$ makes the calculation be non-trivial. 
We demonstrated the explicit calculations in a simple two-flavor model.
The analytic solutions $\chi^{\pm}_{1,2}$ to the equation for the zeros
of the determinant are found.
The large momentum behavier of the solutions is consistent with the
general discussions.
The expression of the thermal effective potential is shown analytically.
The effective potential is completely governed by the solutions to the
equation $\det (p^2 \mathbf{1} + \hat{m}^2 + (\omega \mathbf{1}- i
\hat{\mu})^2) = 0$ but 
extracting the concrete expression of the
solutions is cumbersome when the number of flavors $N$
is large. This fact forces one to utilize the numerical
analysis to find the physics from \eqref{eq:final_result}. 
It is interesting to study applications of our result
\eqref{eq:final_result} by focusing on a specific model.
For example, phase transitions of supersymmetry breaking vacua at early
universe is an interesting topic.
Precise treatments of finite density effects in more realistic 
models are
also interesting.
Apart from the applications, generalizations of our calculational 
scheme to models with vector fields
are important problems.

\subsection*{Acknowledgments}
The work of M. A. is supported by Grant-in-Aid for 
Scientific Research from the Ministry of Education, Culture, 
Sports, Science and Technology, Japan (No.25400280)
and in part by the Research Program MSM6840770029 and 
by the project of International Cooperation ATLAS-CERN of 
the Ministry of Education, Youth and Sports of the Czech Republic.
The work of S.~S. is supported in part by Sasakawa Scientific 
Research Grant from The Japan Science Society and Kitasato University
Research Grant for Young Researchers.

\begin{appendix}
\section{Positivity of $\sqrt{p^2\mathbf{1}+\hat{m}^2}\pm \hat{\mu}$}
 \label{pos}
In this appendix, we prove that 
the matrices $\sqrt{p^2 \mathbf{1} +\hat{m}^2}\pm \hat{\mu}$ are
positive definite when the matrices $\hat{m}$ and $\hat{m}^2 
- \hat{\mu}^2$ are positive definite.
It is sufficient to prove this in the case for $p=0$. 
Suppose that $\hat{m}$ and $\hat{m}^2-\hat{\mu}^2$ are
positive definite, and 
assuming that $\hat{m}-\hat{\mu}$ has a negative eigenvalue, we have
\begin{eqnarray}
 (\hat{m}-\hat{\mu})|a\rangle = a |a\rangle,\qquad a<0, \label{a1}
\end{eqnarray}
where 
$|a \rangle$ is an eigenvector associated with the negative eigenvalue 
$a$. 
By adding the equation (\ref{a1}) multiplied by $\hat{\mu}$, to 
the equation (\ref{a1}) multiplied by $\hat{m}$, we have%
\footnote{
To derive the following equation, we use the
relation $\langle a| [ \hat{m},\hat{\mu}]  |a\rangle =0$
which comes from the fact that $\langle a| A |a\rangle =0 $
for any anti-symmetric matrix $A$ and any real vertor $|a\rangle$.
Since the matrix $\hat{m} - \hat{\mu}$ is real and symmetric,
every eigenvalue of $\hat{m} - \hat{\mu}$ is real
and the corresponding eigenvector is a real vector
except for an overall factor.
}
\begin{eqnarray}
 \langle a|( \hat{m}^2-\hat{\mu}^2)|a\rangle =a \langle a|
(\hat{m}+\hat{\mu})|a\rangle.
\end{eqnarray}
Since the left-hand side is positive by the assumption, 
and $a<0$, we find
\begin{eqnarray}
 \langle a|(\hat{m}+\hat{\mu})|a\rangle<0. \label{a2}
\end{eqnarray}
We now consider the following relation
\begin{eqnarray}
 \langle a|(\hat{m}+\hat{\mu})|a\rangle +  \langle a|
(\hat{m}-\hat{\mu})|a\rangle = 2 \langle a|\hat{m}|a\rangle.
\end{eqnarray}
The right-hand side is positive.
On the other hand, the left-hand side is negative as we have shown in 
\eqref{a1} and \eqref{a2}.
This is a contradiction.
Therefore, we conclude that the matrix $\hat{m}-\hat{\mu}$ does not
have a negative eigenvalue.

If $\hat{m}-\hat{\mu}$ has zero eigenvalue ($a=0$), 
we have 
\begin{eqnarray}
 \langle a|(\hat{m}^2-\hat{\mu}^2)|a\rangle =0,
\end{eqnarray}
which contradicts with the fact that $\hat{m}^2-\hat{\mu}^2$ is positive
definite. 
Consequently, it is proved that
$\hat{m}-\hat{\mu}$ has neither zero nor negative eigenvalues. 
Similarly, one can prove that $\hat{m}+\hat{\mu}$ is positive definite.

\section{Asymptotic behaviors of determinants}\label{large}
In this appendix, we show 
the large-$p$ behavior of the following quantities:
\begin{eqnarray}
&\det(p^2\mathbf{1}+\hat{m}^2+(\omega\mathbf{1}-i\hat{\mu})^2)
\sim{\cal O}(p^{N-4}),&
 \label{p1} 
\\
&{d \over d\omega}\det(p^2\mathbf{1}+\hat{m}^2+(\omega\mathbf{1}
-i\hat{\mu}))\sim{\cal O}(p^{N}). \label{p2}&
\end{eqnarray}
Here we assume that any two of the diagonal elements in $\hat{\mu}$
are not coincident, although, one can reach the same conclusion without 
this assumption.
First we show the relation \eqref{p1}.
We consider the function $f(\eta,\omega)$ in (\ref{eq:eta_equation}). 
We investigate the order in $p$ by expanding $f(\eta,\omega)$ with
respect to $\eta$. 
If we assign an approximate solution $\kappa_k^+(k=1,\cdots N)$ 
to $\omega$ in $f(\eta,\omega)$
, the expansion around $\eta=0$ must start with the order
${\cal O}(\eta^1)$ since the term at ${\cal O}(\eta^0)$ vanishes
from  $f(0,\kappa_i^\pm)=0$.

The matrix in (\ref{eq:eta_equation}) has the following form
\begin{eqnarray}
\left(
 \begin{array}{cccc}
  d_1 & m_{12}^2-\alpha_{12}+\eta\alpha_{12} & \cdots & m_{1N}^2
-\alpha_{1N}+\eta \alpha_{1N} \\
  m_{12}^2+\alpha_{12}-\eta\alpha_{12} & d_2 & & \vdots \\
  \vdots & & \ddots & \vdots \\
  m_{1N}^2+\alpha_{1N}-\eta\alpha_{1N} & \cdots  & \cdots & d_N
 \end{array}
\right),
\end{eqnarray}
where $m_{ij}^2$ and $\alpha_{ij}$ are $(i,j)$-components of $\hat{m}^2$ 
and $[\hat{\mu},\sqrt{p^2\mathbf{1}+\hat{m}^2}]$ respectively.
We have used the properties $m_{ij}=m_{ji}$ and
$\alpha_{ij}=-\alpha_{ji}$. 
For $p \gg 1$, the diagonal component $d_i$
is approximately given by
\begin{eqnarray}
d_i\sim 2p(\mu_i-\mu_k)-\mu_i^2-\mu_k^2+2\mu_i\mu_k+m_{ii}^2-m_{kk}^2
+{\cal O}(1/p).
\end{eqnarray}
Thus $d_i\sim {\cal O}(p)$ for $i\neq k$ and $d_k\sim {\cal O}(1/p)$.
Since the off-diagonal components $m_{ij}^2$
is at order ${\cal O}(1)$,
the highest order term in $p$ includes as many diagonal components
except $d_k$ as possible. 
On the other hand, $\eta$ appears only in off-diagonal components, 
and therefore
non-vanishing terms contain at least one off-diagonal components.
Thus, the terms 
that have the highest power in $p$ should be the following form
\begin{eqnarray}
 && d_1d_2\cdots (m_{ik}^2+\alpha_{ik}-\eta\alpha_{ik})\cdots (m_{ik}^2
-\alpha_{ik}+\eta\alpha_{ik})\cdots d_k \nonumber \\
 && \quad=\left(\prod_{j\neq i,k}d_j\right)((m_{ik}^2)^2-\alpha_{ik}^2
+2\eta\alpha_{ik}^2-\eta^2\alpha_{ik}^2). \label{high}
\end{eqnarray}
For $p\gg 1$, we find
\begin{eqnarray}
 \alpha_{ij}=[\hat{\mu},p+{\hat{m}^2 \over 2p}-{(\hat{m}^2) \over 8p^3}
+\cdots]\sim {1 \over 2p}[\hat{\mu},\hat{m}^2]
 \sim {\cal O}\left({1\over p} \right).
\end{eqnarray}
Substituting them back into (\ref{high}), we find that
\begin{eqnarray}
 {\cal O}(f(\eta,\omega))
 ={\cal O}\left(\left(\prod_{j\neq i,k}d_j\right)(
2\eta\alpha_{ik}^2-\eta^2\alpha_{ik}^2) \right)
 ={\cal O}(p^{N-4}).
\end{eqnarray}

Next we show the relation (\ref{p2}). 
After differentiating $f(\eta,\omega)$ with respect to $\omega$, 
the relation
 $\frac{\partial f(0,\omega)}{\partial \omega}|_{\omega=\kappa_k^\pm}=0$
no longer holds and 
the ${\cal O}(\eta^0)$ term does not vanish. 
Since $\omega$ is only in diagonal components, 
the highest order term in $p$ is obtained when $d_k$ is differentiated. 
From this we find 
\begin{eqnarray}
{\cal O}\left({d \over d\omega}f(\eta,\omega){\Big |}_{\omega=\kappa_k^+}
\right)
={\cal O}\left(d_1d_2\cdots d_{k-1}(2(\omega-i\mu_k))d_{k+1}\cdots d_N
{\Big |}_{\omega=\kappa_k^+} \right)
={\cal O}(p^N).
\end{eqnarray}
\end{appendix}

%%%%%%%%%%%%%%%%%%%%%%%%%%%%%%%%%%%%%%%%%%%%%%%%%

\end{document}